%
%
\magnification 1185
%
%
%
%
 \def\dash{---{}--- } 
 \def\ol{\overline}
 \def\ea{{et al}}
 \def\ie{{i.e.}}
 \def\Par{\par\vskip 6 pt}
 \def\eqn#1{ \eqno({#1}) \qquad }
\newbox\Ancha
\def\gros#1{{\setbox\Ancha=\hbox{$#1$}
   \kern-.025em\copy\Ancha\kern-\wd\Ancha
   \kern.05em\copy\Ancha\kern-\wd\Ancha
   \kern-.025em\raise.0433em\box\Ancha}}
%
 %
 \def\er{{\bf\hat e}_{r}}
 \def\eth{{\bf\hat e}_{\theta}}
 \def\ez{{\bf\hat e}_{z}}

%
\font\bigggfnt=cmr10 scaled \magstep 3
\font\biggfnt=cmr10 scaled \magstep 2
\font\bigfnt=cmr10 scaled \magstep 1
\font\ten=cmr10
%
\leftskip .25 in
\rightskip .25 in
\noindent{\bigggfnt  From circular paths to elliptic orbits: A  } \Par
\noindent{\bigggfnt  geometric approach  to Kepler's motion  } \Par

\vskip 6 pt

 \noindent {\bigfnt
\noindent A Gon\-z\'a\-lez-Vi\-lla\-nueva\dag,
E Guillaum\'{\i}n-Espa\~na\dag,
R P Mart\'{\i}nez-y-Ro\-mero{\dag}\footnote{\ddag}{\noindent\rm On sabbatical leave
from Departamento de F\'{\i}sica, Facultad de Ciencias,\par \vskip -4 pt Universidad
Nacional Aut\'onoma de M\'e\-xico,\par \vskip -4 pt e-mail:
rodolfo@dirac.fciencias.unam.mx},
H N N\'u\~nez-Y\'epez{\S}\footnote{\P}{\noindent\rm On sabbatical leave
from Departamento de F\'{\i}sica, UAM-Iztapalapa,\par \vskip -4 pt e-mail:
nyhn@xanum.uam.mx},
A L Salas-Brito{\dag}\footnote{$^+$}{\noindent\rm Corresponding author,\par \vskip -4 pt e-mail:
asb@hp9000a1.uam.mx or asb@data.net.mx}\Par
}
 \vskip 6 pt

\noindent \dag Laboratorio de Sistemas Din\'amicos, Departamento de Ciencias
B\'asicas, Universidad Aut\'onoma Metropolitana-Azca\-pot\-zal\-co,
Apartado Postal 21-726, C P 04000, Coyoac\'an D F, M\'e\-xico \Par

\noindent $\S$ Instituto de F\'{\i}sica `Luis Rivera Terrazas',
Benem\'erita Universidad Aut\'onoma de Puebla, A\-par\-tado Postal J-48,
C P 72570, Puebla  Pue, M\'e\-xico \Par
 \Par

\vskip 6 pt
\centerline{\bigfnt Abstract}\Par

\noindent The hodograph, \ie\ the path traced by a body in
velocity space,  was introduced by Hamilton in 1846 as an alternative
method for studying certain dynamical problems.  The hodograph of the
Kepler problem was then investigated and shown to be a circle, 
it was next used to investigate some other properties of the motion. 
We here propose a new method for  tracing the hodograph and the corresponding
configuration space orbit in Kepler's problem starting from the  initial 
conditions given and trying to use no more than the methods of synthetic geometry 
in a sort of Newtonian approach. All of our geometric constructions require 
 straight edge and compass only.  \Par

\centerline{\bigfnt Resumen}\Par

\noindent La hod\'ografa, \ie\ la curva recorrida por un cuerpo en el espacio 
de las velocidades,  fu\'e propuesta por Hamilton en 1846 como una alternativa 
para investigar  algunos pro\-blemas din\'amicos. Se demostr\'o entonces que la hod\'ografa 
del problema de Kepler es una circunferencia y posteriormente se la us\'o para
establecer algunas otras propiedades del movimiento. En este trabajo proponemos un 
m\'etodo geom\'etrico semi newtoniano para construir una \'orbita  el\'{\i}ptica  partiendo de sus condiciones iniciales y de la correspondiente hod\'ografa, em\-plean\-do para ello m\'etodos de la geometr\'{\i}a sint\'etica que requieren de la regla y del comp\'as \'unicamente. \Par
\vskip 10 pt

\noindent Classification Numbers: 03.20.+i, 95.10.C\par

\vfill
\eject

\noindent{\bf 1. Introduction}\Par

\noindent   The Kepler problem  has a great deal to offer whenever different approaches to its solution are sought (Rosales \ea\ 1991,  Sivardi\`ere 1992, Mart\'{\i}nez-y-Romero \ea\  1993); this follows since it is a member of the very limited class of superintegrable system (Evans 1990, Mart\'{\i}nez-y-Romero \ea\ 1992,  Salas-Brito \ea\ 1997). One of the simpler approaches to solve the problem  starts by constructing  its hodograph (Gonz\'alez-Villanueva \ea\ 1996, 1998a,b); let us recall that the  hodograph is just  the path traced by the velocity of a body as function of time. In this work we aim to go from  Kepler's hodograph to the corresponding configuration space orbit in a geometric quasi-Newtonian fashion.  Besides the intrinsic beauty of geometrical arguments, we think our approach can contribute to a better understanding of the interrelations between the initial conditions and the  properties of the Kepler problem solutions. \Par

The  hodograph was introduced by Hamilton in the XIX century as an alternative for solving dynamical problems; perhaps the greatest triumph of the method was in the analysis of Kepler motions (Hamilton 1846, Maxwell 1877, Thomson and Tait 1879). However, even if one discovers, as Hamilton did, that the hodograph is circular in such a case, it is natural to wonder how can it be related to the conic section orbits. The problem is  easily solved in the analytical treatment (Gonz\'alez-Villanueva \ea\ 1996, 1998a) but there also exist  beautiful geometric approaches for finding such relation developed by Maxwell (1877) and by Feynman (Goodstein and Goodstein 1996). In this work we want to explain our own version of the geometric relationship between the hodograph  and the orbit, but  we  understand that this approach `cannot substantially differ from what any other [\dots] student can construct' (Chandrasekhar 1995, p xxiii).\Par

\noindent{\bf 2. Kepler's hodograph is circular}\Par

\noindent To establish the circularity of the Kepler hodograph, let us start from the equation of motion of a particle in a fixed Newtonian field at the origin

$$ m{d{\bf v}\over dt}=-{GMm\over r^2} \er,   \eqn{1} $$

\noindent where $m$, $M$, $ G$, ${\bf v}$ and $\er$ are, respectively, the 
mass of the particle, the mass of the attracting body, the Newtonian gravitational constant, the velocity and the unit  radial  vector. As in any central problem, the energy $E$ and the angular momentum 
${\bf L}=m{\bf r}\times {\bf v} =mr^2\dot \theta\ez= L\ez$
are conserved ---the ${\bf L}$-direction defines our $z$-axis.  Please note our convention that the trajectory in configuration space is the {\sl orbit} whereas the trajectory in velocity space is  the {\sl hodograph}. The Kepler orbits are thus confined to a plane orthogonal to ${\bf L}$ which includes the origin; a polar coordinate system with unit vectors $\er$ and $\eth=\ez \times \er$ is chosen in this plane.  Moreover, just from the polar identities $ \dot \er=\dot \theta \eth$, and $\dot\eth=-\dot\theta \er$, it is easy to see that equation (1) can be rewritten as (Moreno 1990)

$$ {\dot {\bf v}} = \enskip {\alpha/m \over r^2} \er    
= \enskip  {\alpha  \over L} \dot \eth,  \eqn{2}$$    

\noindent where we introduced the positive constant $\alpha\equiv GMm$. From (2), it should  be clear that the velocity can be written as (Gonz\'alez-Villanueva \ea\ 1996) 

$$  {\bf v} = {{\bf h}+ {\alpha \over L} \eth } \eqn{3} $$

\noindent where the Hamilton vector ${\bf h}$ is a {\sl constant of motion}. As (3) makes obvious, the velocity traces a circle with radius $R_h=\alpha/L$; the {\sl hodograph}  is  thus  a circle centered  at  ${\bf h}$. Then ${\bf h}$  points along the dynamical symmetry axis of the hodograph---dynamical since it is not only a geometrical property, the interaction intervenes directly; we have found that ${\bf h}$  defines what we call the hodograph's dynamical diameter, the line $\ol{X X_s}$ shown in figure 1.  By extension, equation (3) also shows that every orbit has  a dynamical symmetry axis, which is found geometrically in section 3 below. Notice that $\ol{X X_s}$ can also serve to construct  the notorious Laplace or Runge-Lenz vector ${\gros{\cal A}}\equiv {\bf h}\times{\bf L}$ (Landau and Lifshitz 1976), see section 3 and figure 2,  which points toward the pericentre of the orbit. In what follows, we measure angles in $r$-space counterclockwise  from ${\gros{\cal A}}$, and in $v$-space,  from ${\bf h}$. The circularity of the hodograph implies that the orbit is both periodic and symmetric and that those points on the hodograph where the velocities are antiparallel must be symmetric on the orbit. Notice also that the energy in any Kepler motion can be  related to the  magnitude of  $ {\bf h}$ as follows

$$ E = {m\over 2}\left(h^2 -{\alpha^2 / L^2}\right); \eqn{4} $$ 

\noindent as can be seen in (4), the bounded or unbounded nature of the motions change according to where the $v$-space origin is positioned in relation to the hodograph. It suffices that the $v$-origin be within the hodograph (\ie\ $R_h>h$) to assure that the hodograph is the whole circle and that the orbit is elliptic (Gonz\'a\-lez-Vi\-llanueva \ea\ 1996, 1998a). How to draw the orbit in this case once the hodograph is known is discussed below in section 4.\Par

\noindent{\bf 3. Drawing the hodograph}\Par

\noindent To draw the hodograph given the initial position ${\bf r}_0$ and  velocity ${\bf v}_0$,  we  need to fix the magnitude of the angular momentum $L=m r_0 v_0 \sin\delta$, where $0\leq\delta\leq\pi$ is the angle between the initial position and velocity (see $\delta$ in figure 4); we also need to fix the quantity $R_h=\alpha/L$. Please keep an eye on figure 1 while reading the following.\Par

 Let the point $F$ be the position of the centre of force (hence, the $r$-origin). Draw the line segment $\ol{FR}$ as the
initial position ${\bf r}_0$; extend it up to an arbitrary point $O$
---we are here just choosing the origin in velocity space. From
the $v$-origin $O$, draw the line segment $\ol{OV}$ corresponding
to ${\bf v}_0$ and erect, perpendicular to $\ol{FR}$, a line segment $\ol{OO'}$ of lenght $R_h$ ---that is, we are drawing $-\alpha/L \, \eth$ (recall that
we defined $\eth=\ez\times \er$, where $\ez\equiv {\bf L}/L $). Notice
that the previous construction assumes an attractive interaction. Now sum  
$\ol{OV}$ to  $\ol{OO'}$  to get the point $C$. The line segment $\ol{OC}$, as follows from 
equation (3), represents the Hamilton vector. Having obtained ${\bf h}$, just draw with centre 
at $C$ a circle  of radius $R_h$ to get the hodograph. This geometrical procedure, besides giving ${\bf h}$ and the hodograph, informs us about the  energy  of the motion. It is only a matter 
of noticing whether $O$ is  within the hodograph or
not; if it is within, the energy is negative, if not, the energy
is  positive. Figure 1 illustrates a case in which $O$ is within,
that is, a motion with $E<0$; as follows from equation (4),  $E=0$ occurs when $O$ is 
located precisely on the circle, \ie\ when $h=R_h=\alpha/L$. \Par

 To draw the dynamical symmetry axis of the orbit (\ie\ the line on which the Laplace vector $\gros{\cal A}$ lays) from the given initial conditions, just draw the line segment $\ol{FS}$, which is perpendicular to  $\ol{OC}$ going through the centre of force $F$; as the
segment labeled $\cal{A}$ in figure  2 illustrates. This follows from the paralellism of
${\bf h}$ and the velocity at pericentre ${\bf v}_p$. The line $\ol{FS}$ 
so drawn, {\sl is the  dynamical symmetry axis of the orbit}. Notice also that
${\bf v}_p$ can be drawn by  prolonguing the segment $\ol{OC}$
until it intersects the hodograph. This intercept is marked $X$ in
figure 1. If, as happens in figures 1, 2, and 3, there are two intersections with
the hodograph and not just one, the velocity space origin  $O$ is  necessarily inside the hodograph, that is, it always corresponds to the case $E<0$. The second intercept, labeled $X_s$ 
in figure 1, defines the segment $\ol{OX_s}$ corresponding to the velocity at the 
{\sl apocentre} of the orbit, \ie, at the point on the orbit farthest from the
centre of force where the speed is the lowest possible.\Par

\noindent {\bf 4.  How to trace the elliptic orbit}\Par

\noindent Let us assume that the origin of coordinates in velocity space
happens to be inside the circle of the hodograph; this is the case whose 
realization from initial conditions was discussed in section 2 and illustrated in figure 1.
Please refer to figure 2 for the schematic representation of the geometric 
steps that follow.  The points $F$, $R$, $O$, $O'$, $V$ and $C$ in
figure 2 have exactly the same meaning as in figure 1, that is, they
serve to construct  the Hamilton vector $\ol{OC}$ and the hodograph
centered at $C$ given the initial conditions ${\bf r}_0$ (the straight
line $\ol{FR}$, which makes an angle $\theta$ with ${\cal A}$), and ${\bf v}_0$ (the straight line $\ol{OV}$) and the vector $-\eth R_h $ (the straight line $\ol{OO'}$). The initial velocity also helps to define  the segment $\ol{CV}$  making the same angle $\theta$ with ${\bf h}$. In fact, we will always assume this meaning for the labeling of points in figures 2--4. \Par

To locate any point on the  orbit, first extend the straight line
$\ol{OV}$ until it again intercepts the hodograph at  $V_s$
(see figure 2).  Then trace a perpendicular to $\ol{CV_s}$ passing through
 $R$,  this line intercepts the symmetry axis (drawn as
in section 3) at the auxiliar point $F'$. To
locate the point on the orbit corresponding to any given point on the
hodograph, let us first notice that we have already one such pair of
points, the initial conditions: point $R$ and point $V$. Let us choose
another point $V'$ on the hodograph, then draw the straight line
$\ol{OV'}$ and  extend it until it intersects the hodograph at point
$V_s'$. Draw  two straight lines perpendicular to $\ol{CV'}$ and to
$\ol{CV_s'}$ passing through $F$ and $F'$, respectively; we assert that
these two perpendiculars meet at the required point $R'$ on the orbit,
as was the case with the perpendiculars to the  segments
$\ol{CV}$ and $\ol{CV_s}$, related to the initial conditions and meeting
at $R$. To draw the complete orbit, \ie\ the gray curve in figure 3, we 
have to repeat the  procedure starting from each point on the 
hodograph. \Par

To decide the shape of the constructed orbit, 
 draw the circular arc ${F'W}$  centered at $R$ with a radius
equal to the lenght of $\ol{F'R}$; this arc intercepts the straight line
$\ol{FO}$ at the point $W$ (see figure 3).  Next, trace the circular arc
$WW'$ centered at $F$ with radius $\ol{FW}$. It is now easy to see, just
by noticing that the  shaded triangles $\triangle V'V_s'C$ and
$\triangle W'F'R'$ are both isosceles and similar to each other,  
that the point $R'$ on the orbit is at the
same distance from the point $W'$ than from the point $F'$. We can see
thus that the radius of the circular arc $WW'$ is the sum of the lenghts
of $\ol{FR'}$ and $\ol{F'R'}$ and, therefore, that {\sl in the case
$E<0$ the orbit is necessarily an ellipse} whose major axis $2a$ equals
the lenght of the line  $\ol{FW}$. The auxiliary point $F'$ is
thus seen to be the second focus of the ellipse, the first one coinciding with the
centre of force $F$.  The line $\ol{FS}$  is parallel to the
symmetry axis of the ellipse as we had anticipated. In fact, the
eccentricity of the ellipse is easily calculated as
$\epsilon= h/R_h=OC/CV $ (Gonz\'alez-Villanueva \ea\ 1996). Thus,
 ${\gros{\cal A}}$, is the line segment parallel to $\ol{FS}$ of lenght 
$\alpha\epsilon$ pointing towards the pericentre. The construction performed here also shows that $\ol{FR'}$ always makes the same angle with $\ol{AB}$ than $\ol{CV'}$ with $\ol{OC}$ (figures 2 and 3). Geometric methods for reconstructing the orbit in the cases $E\ge0$  are described 
in detail in (Gonz\'alez-Villanueva \ea\ 1998c). \Par

\noindent {\bf 5. Why the method works}\Par

\noindent Let us first pinpoint the uniqueness of the elliptic orbit drawn in figure 3; this follows  since the initial conditions, ${\bf r}_0$, ${\bf v}_0$, uniquely especify both ${\bf L}$ and ${\bf h}$. These, in turn, are the necessary and sufficient conditions to obtain ${\gros{\cal A}}$  and, hence, both the dynamical symmetry axis and the orbit (Gonz\'alez-Villanueva \ea\ 1996, 1998b). But, even with the uniqueness established, the relationship of the hodograph with  the orbit, and the lines used in section 3, can still remain obscure. How can we ascertain that the velocity at any point on the hodograph is parallel to the tangent at the corresponding point on the ellipse? An attempt to explain the situation is in  figure 4, which re-elaborates figure 3 removing certain  unnecesary features, and in the explanation that follows: \Par

Let us assume that the construction of  section 3 has been carried out. To start the explanation, draw a circle (see figures 3 and 4) with radius  $\ol{FW}$ (\ie\ radius $2a$) centered at $F$ (this corresponds to the arc $WW'W''$ in figure 3); trace the straight line $\ol{AB}$, corresponding to the  ellipse's dynamical symmetry axis, on this line mark the second focus $F'$.  Next, pick an arbitrary  point $R$ on the orbit and trace the   segment $\ol{FR}$ making an angle $\theta$ with $\ol{AB}$ and extend it until the intercept $W$ on the circle. This defines the  segments $\ol{F'W}$ and its continuation  $\ol{F'W_s}$. Now trace their perpendicular bisectors $\ol{MR}$ and $\ol{M_sR_s}$, these lines intercept  $\ol{FW}$ and $\ol{FW_s}$ at  $R$ and $R_s$, respectively.  Extend $\ol{FW_s}$ until it intercepts the circle at $W''$ and trace the lines $\ol{FW''}$ and $\ol{WW''}$. Thence the triangles $\triangle FWW''$, $\triangle RWF'$ and $\triangle R_sF'W_s$, are  isosceles, simililar to each other and with common angles $\delta$ and $\pi-2\delta$. From this, it is easy to see that $\ol{W_sR_s} = \ol{R_sF'}$ and $\ol{F'R} = \ol{RW}$, \ie\ they belong to the same ellipse. Besides, the lines $\ol{F'R}$ and $\ol{FR_s}$ are parallel by construction, the same is true of lines $\ol{R_sF'}$ and $\ol{FR}$, therefore $\ol{FR_s}$ and $\ol{F'R}$ make the same angle $\xi$ with $\ol{AB}$. \Par

 Notice that $\triangle WW_sW''$ is a right triangle by construction thence, $\ol{WW''}$ and the perpendicular bisectors $\ol{MR}$ and $\ol{M_sR_s}$ are  parallell to each other. Notice that the perpendicular bisector to segment $\ol{WW_s}$ also bisects the angle $\angle WFW_s$ and therefore that the lines $\ol{MR}$ and $\ol{M_sR_s}$ are tangent to the ellipse at $R$ and at $R_s$. This establishes that, in effect, the tangent at every point on the orbit is parallel to the {\sl corresponding} velocity on the hodograph and, at the same time, that  every pair of symmetric points on the hodograph, {\sl where the velocities are antiparallel}, corresponds to a pair of symmetric points on the orbit. \Par

 It is now easy to see that if we rescale the circle $WW'W''$ in figure 3, or $AWW''$ in figure 4,  by the factor $\alpha/2aL$ we get essentially the hodograph but rotated $\pi/2$ respect that in figure 1;  furthermore, under these same rescaling and rotation, the lines $\ol{F'F}$, $\ol{F'W}$ and $\ol{F'W_s}$ in figure 4, become, respectively, the Hamilton vector and the velocities at $V$ and $V_s$, all shown in figure 3.\Par

\noindent {\bf 6. Concluding remarks}\Par

\noindent We have shown how the bounded orbits of the Kepler problem can
be drawn starting from the initial conditions---and the hodograph---using no more
than straight edge, compass and a few lines in a piece of paper. We have also exhibitted that  the hodograph and ${\bf h}$ are crucial for deciding {\sl geometrically} if the orbits are bounded or not and, furthermore, that with their help, we can draw any  orbit starting from
arbitrary initial conditions. Although we have not shown it here, these elementary geometrical techniques can be quite useful for discussing  orbital manoeuvres and other features
of the motion in a Newtonian field. This means that our approach can 
provide a very convenient method for addressing the interplay between the physics 
and the geometry of Kepler's problem in a kind of Newtonian fashion. \Par

The need to make the construction presented more accesible has prompted us to program our 
construction of the orbits using  The Geometer's
Sketch\-pad3$^{\hbox{\copyright}}$ a very nice piece of software for doing
 geometrical constructions which can be obtained at {\sl http://www.keypress.com\-/product\_info/sketch-demo.html} in a
demo version. The demonstration of our constructions has been succesful 
with the students. Any interested reader may try to reproduce the method using the simple language associated with the Sketchpad. \Par

As a final remark, we must say that our main motivation for this work must be found on the amusement side. We  have had a lot of fun in trying to do mechanics without using most of the usual analytic techniques of contemporary physics. We hope this article may convey to the readers the sense of 
enjoyment we discovered in the geometric beauty of dynamics. In our eyes at least---though beauty is in the eyes of the beholder!--- these considerations are enough to justify the quasi-Newtonian approach presented in this article. To finalise, we found convenient to  paraphrase Chandrasekhar's paraphrasing of Ben Johnson (Chandrasekhar 1995, Epilogue) since it clearly summarizes our viewpoint: {\sl Newton's methods were not of an age, but for all time!}  \Par
\vfil
\eject

\noindent{\bf Acknowledgements}\Par

\noindent  This work has been partially supported by CONACyT (grant
1343P-E9607).  We want to thank E Pi\~na-Garza for inspiration; we also wish him a quick and complete recovery. We thank  our colleagues L Fuchs-G\'omez, R G\'omez, D Moreno, K Quiti, and J L del-R\'{\i}o-Correa for their useful comments and/or advice.  Thanks are also due to members of the Taller de Matem\'aticas of the Facultad de Ciencias-UNAM for sharing with us their knowledge of the  The Geometer's Sketchpad3$^{\hbox{\copyright}}$. This work is dedicated to the memory of A  P Pardo, M Mina, Q Motita,   M Kuro, B Minina, M Miztli, M Tlahui and B Kot. Last but not least AGV wants to express his warmest thanks to  Armida de la Vara and Luis Gonz\'alez y Gonz\'alez  for all the  support and encouragement of the last 12 years. \Par

\vskip 15 pt
\vfil
\eject

\noindent{\bf References} \Par

{\ten \baselineskip 11.5 pt

 \noindent Chandrasekhar S 1995 {\it Newton's Principia for the common
reader} (Oxford: Clarendon Press) \Par

 \noindent Evans N W  1990 {\it Phys.\ Rev.\ } A {\bf 41} 5666 \Par

\noindent  Gonz\'alez-Villanueva A,  N\'u\~nez-Y\'epez H N and
Salas-Brito A L, 1996  {\it Eur.\ J.\ Phys.\ } {\bf 17} 168  \Par

\noindent \dash 1998a 
{\it Rev.\ Mex.\ Fis.\ } {\bf 44} (1998) 183\Par

\noindent  Gonz\'alez-Villanueva A,  Guillaum\'{\i}n-Espa\~na E,
N\'u\~nez-Y\'epez H N and  Salas-Brito A L,  1998b {\it Rev.\ Mex.\ Fis.\ } 
{\bf 44} (1998) 380\Par

\noindent  Gonz\'alez-Villanueva A,  Guillaum\'{\i}n-Espa\~na E, Mart\'{\i}nez-y-Romero R P,
N\'u\-\~nez-Y\'e\-pez H N and  Salas-Brito A L,  1998c FC-UNAM preprint\Par

\noindent  Goodstein D L and  Goodstein J R, 1996 {\it Feynman's lost
lecture. The motion of planets around the sun} (New York: Norton) Ch 4\Par

\noindent Landau L and  Lifshitz E M 1976 {\it Mechanics} (Oxford:
Pergamon)\Par

 \noindent Hamilton W R 1846 {\it Proc.\ Roy.\ Irish Acad.\ } {\bf 3} 344 \Par

  \noindent Mart\'{\i}nez-y-Romero R P, N\'u\~nez-Y\'epez H N, and
Salas-Brito A L  1992 {\it Eur.\ J.\ Phys.\ } {\bf 13} 26 \Par

  \noindent \dash  1993 {\it Eur.\ J.\ Phys.\ } {\bf 14}
1--3 \Par

\noindent Maxwell J C 1877 {\it Matter and motion} 1952 reprint (New York:
Dover) 107 \Par

\noindent Moreno D 1990  {\it Gravitaci\'on Newtoniana} (M\'exico City:
FCUNAM)
 \Par

\noindent Rosales M A, del-R\'{\i}o-Correa J L, Castro-Quilant\'an J L
1991 {\it Rev.\ Mex.\ Fis.\ } {\bf 37} 349 \Par

\noindent Salas-Brito A L, Mart\'{\i}nez-y-Romero  R P, N\'u\~nez-Y\'epez
H N 1997  {\it Intl.\ J.\ Mod.\ Phys.\ A} {\bf 12} 271  \Par

\noindent Sivardi\`ere J 1992  {\it Eur.\ J.\ Phys.\ } {\bf 13} 64\Par

\noindent  Thomson W and Tait P G 1879 {\it Treatise on natural philosophy}
 1962 reprint (New York: Dover) \S 37--\S38 \Par
}
 \vfill
 \eject

\centerline{\biggfnt Figure Captions}\Par

\noindent Figure 1\par

\noindent The geometrical procedure for obtaining both the Hamilton vector
and the hodograph from given initial conditions ${\bf r}_0$ and ${\bf v}_0$
is illustrated. $O$ labels the origin of coordinates in velocity space or
$v$-origin and $F$ labels the location of the centre of force or $r$-origin. To draw the
segment ${OO'}$, corresponding to $-\eth\alpha/L$, we assumed that
${\bf L}$ points outside the plane of the paper. The Hamilton vector is the
line segment $\ol{OC}$, the circle  $X_sVX$ centered at $C$ is the
hodograph. The straight  segments $\ol{SF}$ and $\ol{SX}$ correspond,
respectively,  to the dynamical symmetry axes of the orbit and of the
hodograph. The discussion related to this figure can be found in section
3.\Par
\vskip 5 pt

\noindent Figure 2\par

 \noindent The procedure for reconstructing the orbit when the hodograph
encompass the $v$-origin is illustrated. $F$,  corresponds to the position of the centre of force,  $O$ is the $v$-origin, $V$ and $R$  are an arbitrary velocity on the hodograph and its corresponding position on the elliptic orbit, and $C$ is the geometric centre of the hodograph. ${\cal A}$ represents the Laplace (or Runge-Lenz) vector. For a detailed discussion  of the method for reconstructing the orbit see section 4. \Par

\noindent Figure 3\par
 \noindent To prove that the orbit is indeed an ellipse (the only case considered in this article), we need the help of an auxiliary circle with radius $2a$, equal to the lenght of $\ol{W_sW''}$, and to recognize that the two shaded isosceles triangles $\triangle R'W'F'$ and $
\triangle CV'V_s'$ are similar to each other. \Par

\vskip 5 pt

\noindent Figure 4\par

\noindent This is essentially figure 3 excepting for some  details unnecessary  for the explanation in section 5. The purpose of this figure is to explain the  reasoning behind the method used to construct the orbit starting from the hodograph. The circle $AWW''$ (with radius $2a$) corresponds to the circle $WW'W''$ in figure 3 and, after a $\pi/2$-rotation and a rescaling by $\alpha/2aL$, it  also corresponds to the hodograph in figures 1, 2 and 3.  It is convenient to remember that a conic  can be defined as the locus of points  being at the same distance from both a fixed point ($F'$) and  a fixed circle (the arc $WW'W''$).  The angle between the initial position and the initial velocity is $\delta$.\Par

 \vfil
 \eject
 \end